\documentclass[smallextended]{svjour3}
\smartqed

\usepackage{makeidx}  
\usepackage{graphicx}
\usepackage{subfigure}
\usepackage{listings}
\usepackage{mathrsfs}
\usepackage{textcomp}
\usepackage{multirow}
\usepackage{amssymb}
\usepackage{natbib}
\usepackage{color}

\usepackage[breaklinks=true]{hyperref}
\journalname{Ambient Intelligence and Humanized Computing}
\begin{document}

\title{Secure Ambient Intelligence Prototype for Airports
}
\titlerunning{Secure Ambient Intelligence Prototype for Airports
}  
%
\author{Nayra Rodr\'iguez-P\'erez \and Josu\'e Toledo-Castro \and Pino Caballero-Gil \and Iv\'an Santos-Gonz\'alez \and Candelaria Hern\'andez-Goya}
\institute{Departamento de Ingenier\'ia Inform\'atica y de Sistemas, \\Universidad de La Laguna, Tenerife. Espa\~na\\
\email{\{mrodripe, jtoledoc, pcaballe, jsantosg, mchgoya\}@ull.edu.es}
}

\date{Received: date / Accepted: date}

\maketitle

\begin{abstract}
Nowadays, many technological advances applied to the Internet of Things (IoT) make the introduction of innovative sensors aimed to deploy efficient wireless sensor networks possible. In order to improve the environment and people’s lives, real time analysis of certain environmental variables may favour the reduction of health risks related to the deterioration of air quality. To this respect,  the proposed system implements a particular prototype of IoT device characterized by the assembly of ambient sensors capable of measuring pollutant gases, temperature and humidity. For this purpose, Raspberry Pi and Arduino platforms are used. Several security methods are introduced to ensure the integrity of air quality data by implementing Merkle Trees on each IoT node and on the Cloud server. Besides, the authenticity of IoT devices and the confidentiality of communications are guaranteed by implementing HTTPS requests. Finally, authentication tokens are used to identify system users, and different security rules are applied to manage database operations.
	
	\keywords{Air Quality, Wireless Sensor Networks, Merkle Trees, Security, Airports}
\end{abstract}

\section{Introduction}

Poor indoor air quality in buildings and other infrastructures can cause health problems for people in the short and long term. Exposure to certain pollutant gases can produce immediate effects such as irritation of the eyes, nose, headaches or fatigue, among many others \cite{brown2019indoor}. In addition, certain respiratory problems such as asthma can be significantly aggravated as a result of exposure to certain polluting gases \cite{trasande2005role}.

Several gases and pollutants, such as carbon monoxide (CO), carbon dioxide (CO2), sulphur dioxide (SO2), particles (PM2.5 and PM10) and nitrogen oxides (NOx), directly condition the air quality of indoor places. All these variables are commonly used for air analysis and the construction of air quality indices \cite{bernstein2008health}. Likewise, certain environmental conditions such as temperature and humidity affect the concentration level of the above gases as well as the performance of certain sensors currently available for their measurement  \cite{fang1998impact}. For this reason, they must be taken into account when developing a model for the study and analysis of the environment and air quality in  indoor areas.

A continuous monitoring  of these pollutants in real time is essential in public transport infrastructures such as airports. The air quality in these structures can affect the health of a large number of passengers in transit, so its control is a priority and the efficient implementation of monitoring is a challenge to be resolved to ensure public health \cite{schlenker2015airports}. 

Airports are one of the biggest sources of air pollution in many cities in the world. The number of people in transit at airports and terminals is very high. For example, Spain is among the first countries in Europe in air traffic, with  263,7 million  passengers in 2018 \cite{aena:Online}. Pollution is a major threat to the health of airport users due to its prolonged stay indoors with closed environments and especially for airport staff. The standardization of air quality protocols and their application  indoors, especially at airport terminals, is important to ensure the wellness of the people in transit.

New technologies and ambient intelligence  may be important elements for efficient solutions to the aforementioned problem related to the health of people. In order to respond to this need, the deployment and integration of devices related to the Internet of Things (IoT) \cite{atzori2010internet} and wireless sensors across the environment is  proposed here to provide an air quality monitoring service. Thus, different technologies and applications that deploy Wireless Sensor Networks (WSN) \cite{akyildiz2002wireless} to provide innovative solutions are used.

The system proposed in this paper is based on the integration of WSN aimed at providing a prototype capable of monitoring the air quality within  airport terminals. To this respect, the objective is to perform a preventive model based on analysing the air pollution status in real time across certain airports' indoor areas of interest. It is also intended to provide a secure and efficient system that ensures the integrity, confidentiality and authenticity of environmental data measured by IoT devices and sensors.

This paper is organized as follows. Some related works are detailed in Section 2. Section 3 includes the main details of the proposal. The  method applied to study the air quality information is explained in Section 4. Then, Section 5 details the security methods applied to ensure the environmental data integrity by introducing Merkle trees, and to highlight other integrated system security practices. Finally, Section 6 provides some conclusions and future work.

\section{Related work}
\label{sec:relatedwork}

Nowadays, particular attention is paid to the study, analysis and monitoring of the air quality of indoor and outdoor areas, aimed at avoiding risks for  people's health. Firstly, a study of environmental monitoring systems with sensor networks has been carried out. There are several studies that deal with this topic by applying different methodologies and control methods based on implementing wireless sensor networks. In the work  \cite{morawska2018applications}, the monitoring of air quality is studied by several research teams to analyse the possibilities provided by low-cost technologies. In this way, this work makes it possible to study the reliability of applying low-cost sensor technologies and networks for air quality monitoring. Another example of environmental monitoring is detailed in the work  \cite{sipani2018wireless}, which proposes the monitoring and control of environmental factors through  low-cost boards and sensors. In this case, only  temperature and humidity are monitored, activating a cooling system to control the temperature levels. Similarly, the work \cite{lombardo2018wireless} studies the distribution of small wireless sensor nodes aimed to monitor different environmental parameters (temperature and humidity) and store the data at different levels.

Regarding environmental intelligence, several works use this methodology to try to  improve people's lives. A more extended definition of its methodology and use has been studied at work \cite{aarts2009ambient}, highlighting the application of ambient intelligence for monitoring and emergency management. Through applying various mechanisms and distributing sensors at home, an environment for home cognitive rehabilitation is proposed in the paper \cite{oliver2018ambient}. To this respect, the use of ambient intelligence systems is highlighted with the aim of using technologies capable of receiving data from the user's environment. As a result, actions are performed on the basis of this obtained information. Several use cases based on ambient intelligence are presented in \cite{tapia2010agents}. Where the life cycle of an ambient intelligence system is detailed, the importance of analysing sensor data is highlighted and different examples of distributed services and applications are shown.

The application of security layers is an important area in  systems related to the IoT. In this regard, the use of Cloud Computing and IoT technologies together with the analysis of general security problems are highlighted in the paper \cite{stergiou2018secure}. The advantages of integrating both technologies and their common characteristics are shown here. The work \cite{przydatek2003sia} proposes a framework in which secure information is added in large sensor networks to avoid physical vulnerabilities, in addition to allowing the user to verify a given response. A study of public-key cryptographic protocols has been carried out here and particular attention has been paid to  Merkle trees \cite{merkle1980protocols}. With regard to this, the work \cite{mao2017position} proposes a publicly  verifiable scheme using Merkle trees aimed to guarantee the protection of the integrity of the cloud data. In this work, a tree  in which each node knows the position of its primary nodes instead of the whole tree for the verification of integrity is proposed. The authors of \cite{li2014efficient} propose an authentication scheme for smart grid communication based on using efficient Merkle trees. The purpose is to avoid relevant critical security challenges in smart grids, such as message injection attacks or replay attacks.

The system proposed in this document unifies the subject matter previously studied in an independent manner. This system is based on the use of the Internet of things for the monitoring of indoor air quality by applying ambient intelligence. A monitoring system of certain environmental variables, relevant in the analysis of air quality, is developed. Statistical calculations are performed by the system to analyse collected air quality information. For this purpose, WSNs are implemented on the basis of a particular IoT prototype that is distributed across critical transport infrastructures. The aim is to monitor several pollutant gases and other environmental conditions, such as temperature and humidity. In addition to this, a security layer is applied to provide secure communications and ensure the integrity of air quality information by implementing Merkle trees on each distributed IoT device. Providing a layer of security in these devices is a fundamental aspect in this system.

\section{Proposed system}
\label{sec:system}

The  developed system is composed of three main components. Firstly, there is a prototype of IoT device assembled with different environmental sensors that are relevant to perform real time monitoring of air quality. These sensor nodes are distributed in different indoor areas of certain critical transport infrastructures, such as airports. To this respect, the aim is to monitor in real time the state and variations in air quality in user transit zones. Several variables are measured in real time, such as pollutant gases and other important ambient conditions (temperature and relative humidity). Regarding polluting gases, carbon monoxide (CO), carbon dioxide (CO2), nitrogen oxides (NOx), sulfur dioxide (SO2) and the presence of smoke are used. In addition to this, temperature and humidity are monitored because their values affect the state of the aforementioned variables and the efficient measurement process of each assembled sensor. This proposal analyses several gas pollutant variables very relevant in the field of air quality monitoring. The implemented prototype of IoT device can be improved by assembling other compatible sensors with the aim of monitoring different environmental variables.

\begin{figure}
	\centering
	\includegraphics[width=\textwidth]{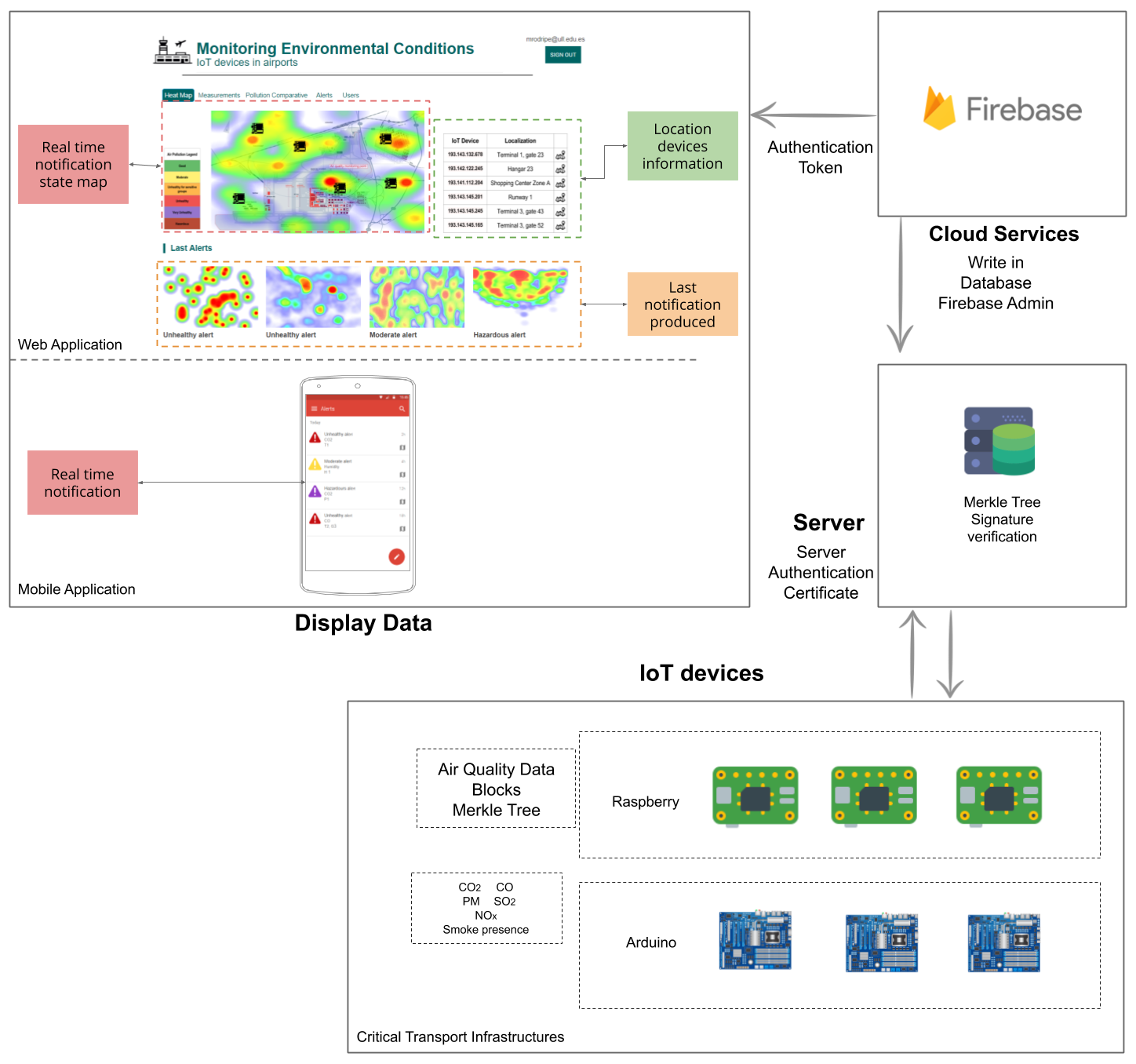}
	\caption{General Scheme}
	\label{fig:nuevoEsquema}
\end{figure}

In the prototype, each measurement cycle is performed every 10 minutes, although the measurement frequency of each IoT device can be increased if the current air quality state requires it. Once gas pollutant concentrations and ambient conditions are measured, the values are digested by the SHA-3 hash function and added to a Merkle tree as a new leaf node, as explained below. Every distributed IoT device implements a Merkle tree to ensure the integrity of all measured air quality data blocks . Every air quality data block is composed of a value of every monitored gas pollutant and a measurement of temperature and relative humidity.

Each WSN master node (considered as gateway with regard to the proposed topology network) communicates wirelessly via Wifi with the Cloud server and via HTTPS requests. Thus, the client certificate corresponding to every IoT device is attached. For each request made by an IoT device, the following data is sent to the server:

\begin{enumerate}
	\item Ambient measurements. Concentrations of monitored pollutant gases (CO2, CO, SO2, NO2, etc.), temperature value and humidity percentage.
	\item Data required to verify the integrity of the Merkle tree built up to then: root node and intermediate non-leaf nodes required to build and verify that root.
	\item Device configuration and status parameters such as battery status, MAC address, etc.
	\item Client certificate to allow the authentication of the corresponding IoT device in the Cloud server.
\end{enumerate}

The Cloud server was  implemented with Nodejs and Express \cite{nodejs2019:Online} \cite{express2019:Online} and was  launched over HTTPS, following the MVC pattern (Model, View, Controller) \cite{leff2001web}. The mobile application was  developed with Android \cite{android2019:Online} and  the web application was  implemented with Vuejs \cite{vuejs2019:Online}.

The notation used to denote all elements in the system is shown in table \ref{tab:notation}.

\begin{table}[]
	\centering
	\caption{Proposed Notation}
	\label{tab:notation}
	\begin{tabular}{|c|c|c|c|c|c|c|}
		\hline
		Notation        &   Value                  \\ \hline
		CO2             &   Carbon Dioxide         \\ \hline
		CO              &   Carbon Monoxide        \\ \hline
		NOx             &   Nitrogen Oxide       \\ \hline
		SO2             &   Sulfur Dioxide         \\ \hline
		PM              &   Particulate matter     \\ \hline
		T               &   Temperature            \\ \hline
		H               &   Humidity               \\ \hline
	\end{tabular}
\end{table}

\subsection{Air quality notifications}

Each time a new environmental block on air quality is sent to the server, the integrity and confidentiality of air quality measurements are checked on the basis of analysing the implemented Merkle tree. Likewise, IoT device authenticity is verified by checking its client certificate. Then, the new air quality data block is analysed and stored in the database. For each air quality measurement (CO2, CO, NOx, etc.), several statistical calculations are made to interpret the current state, variations and the dispersion of measurements over time (detailed in Section \ref{sec:method}). Once the statistical results are obtained, environmental status notifications are activated and sent to the system users through different platforms: Web service, Firebase and the implemented mobile application.

Environmental status notifications are generally sent every 30 minutes when the air quality does not pose any risk to the health of users and maintains its state over time so that the analysis of air quality returns good results. On the other hand, notifications are considered air quality alerts when the statistical evaluation of air quality data blocks in a particular indoor monitored area provides results that may cause health risks. In contrast, air quality alerts are sent every 15 minutes, although the  frequency of shipment may change depending on air quality measurements. For this purpose, the average and the standard deviation of every monitored variable are analysed to evaluate the dispersion and radical changes in the environmental variable.

The following fields are defined for a new environmental status notification or alert with regard to a particular environmental variable (polluting gas, temperature or humidity):

\begin{enumerate}
	\item Ambient variable that has caused the air quality alert and the value of its last measurement.
	\item Ambient variable state (good, moderate, bad).
	\item Average of the ambient variable calculated using the last 50 measurements registered.
	\item Coefficient of variation with respect to the average value (percentage).
	\item Color representation of the current state of the environmental variable.
	\item Affected indoor zone. Area delimited by the IoT device that measures the environmental variable values that has provided a new alert.
\end{enumerate}

Alerts generated after analysing environmental measurements are displayed in the system through a color code that indicates their status:
\begin{itemize}
	\item Green: This color is used when air quality is considered satisfactory and air pollution does not involve any health risk.
	\item Yellow: Air quality is moderate, even though for a few people there may be a minor health  problem due to pollutants.
	\item Red: Health alert, the entire population is more likely to be affected. In this case, an alert notification is generated to warn about the existence of emergency conditions.
\end{itemize}

Alerts are sent to users through the Firebase platform, which is a BaaS system (Backend as a Service). This platform provides the following services: NoSQL realtime database (Firestore service), authentication (Firebase Authentication), push notifications (Cloud Notification Push) and backend functions executed on the cloud (Cloud Functions). Each new air quality alert generated and stored in the database triggers an implemented cloud function in charge of sending a new notification to the system users. This notification requires an identification token from every mobile device. This token is generated during the installation of the mobile application.

\subsection{IoT prototype}

A Raspberry Pi Model B+ device has been considered to manage a group of Arduino-based sensor nodes capable of performing environmental measurements in particular indoor areas. The aim is to monitor the air quality over areas of interest in critical transport infrastructures.
In comparison with other IoT platforms, the Raspberry Pi allows wireless connectivity through Bluetooth 4.2 BLE and Wifi 802.11ac (2.4GHz / 5GHz). These devices provide a greater computational capacity and memory that favour the management of Merkle trees to ensure the integrity of all measured air quality data blocks. A Wifi module assembled in the Raspberry Pi's mother shield is used to send every new air quality data block over HTTPS requests to the developed Cloud server. With regard to a master-slave topology network, Raspberry Pi device is considered the master device of each sensor network based on Arduino devices, so thus is the gateway in charge of receiving and processing the environmental data packages measured by Arduino nodes. With respect to this, several Arduino microcontrollers are distributed as slave sensor nodes and coordinated by a master device to perform environmental measurements.

Thus, several sensor nodes based on Arduino Mega's shield are distributed to perform environmental measurements of gas concentrations, temperature and humidity at indoor areas close to their master Raspberry Pi device. This proposal of WSN is proposed to improve the analysis and classification of the air quality data related to every monitored area of interest jointly by extending the air quality measurement area managed by every master device (cluster) \cite{Caballero}. Arduino - Raspberry Pi communications are made by assembling an HC-05 Bluetooth module in the microcontroller and using the Bluetooth module built-in by default in the Raspberry-Pi node. The following points explain in detail the implemented environmental cycle in detail: 

\begin{enumerate}
	\item The air quality of a particular area in a critical transport infrastructure is required to be monitored, so a WSN composed of a master Raspberry Pi device and several Arduino-based slave nodes are distributed.
	\item Polluting gases, temperature and humidity are successively measured by specific environmental sensors assembled to the Arduino shield. Thus, analogue and digital readings are performed to get input data of polluting gas concentrations and ambient conditions (temperature and humidity), respectively.
	\item The data obtained by every measurement cycle is sent to the master Raspberry Pi device by the aforementioned HC-05 module.
	\item The master IoT device receives the data and adds it to a Merkle tree that is constantly updated for each new data block of air quality measurements. This tree ensures the integrity of all its air quality data and is continuously verified by the Cloud server.
\end{enumerate}

\begin{figure}
	\centering
	\includegraphics[width=\textwidth]{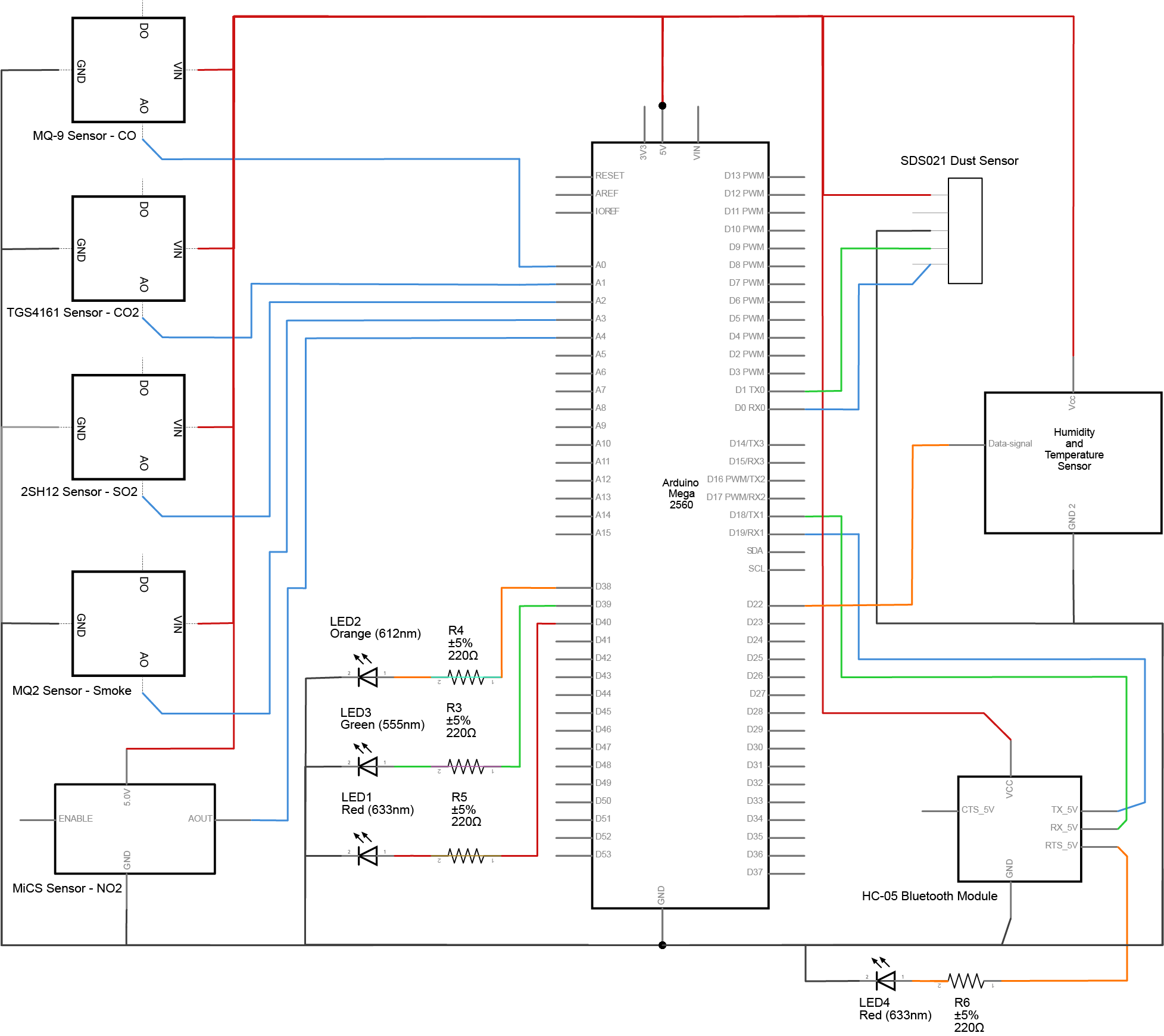}
	\caption{Arduino-based IoT prototype (slave node)}
	\label{fig:IoTPrototype}
\end{figure}

The proposed wireless sensor network may change its distribution and composition depending on the size of the indoor area that is pretended to be monitored. Certain aspects are considered to manage the implemented WSN, such as ensuring availability of communications through Bluetooth modules. With regard to this, sensor networks may require many Arduino-based sensor nodes (considered as slaves) when it is necessary to cover  large indoor areas. However, other sensor networks aimed to monitor small-sized zones in critical transport infrastructures may only need a sensor node. In such cases, the master Raspberry Pi device measures the analysed environmental variables by assembling the same digital / analog sensors that are integrated into Arduino-based sensor nodes. As a consequence of the digital pinout disposed on Raspberry Pi’s shield, the MCP3008 \cite{mcp3008:Online} analogue-digital converter has been assembled on it. Its aim is to enable eight analogue channels that allow the Raspberry Pi’s master device to measure analogically the gas polluting concentrations. The rest of the sensors shown in Figure \ref{fig:IoTPrototype} to measure particles in suspension and atmospheric variables, which means temperature and humidity, are digitally measured by the pinout of the Raspberry Pi’s shield.

Regarding the power supply for Raspberry Pi's pinout, voltage converters have also been  integrated to avoid damage to the GPIO (General Purpose Input/Output) pinout. These pinouts work with 3.3V instead of 5V. Several sensors return voltage results of gas concentrations close to 5 V, so it is necessary to scale them by taking a maximum voltage of 3.3 V.

\subsection{Air quality sensors}

Table \ref{tab:environmental_specification} shows the list of the proposed environmental sensors to be assembled with the IoT device motherboard. Each of them allows the monitoring of a different environmental variable, which means a particular polluting gas or environmental condition (temperature and humidity). The efficiency of the measuring process of the variable performed by each sensor depends on different environmental conditions of temperature and relative humidity. Most of the proposed sensors require 5V (volts) for power supply. Analogue readings have been implemented in order to register proportional measurements of each polluting gas concentration and the presence of particles (PM2.5 and PM10). For this purpose, analogue pinouts of Arduino shields have been used. If the Raspberry Pi device is in charge of measuring polluting gases, the aforementioned analogue-digital converters are required. However, dust particles and ambient conditions of temperature and humidity are digitally measured with SDS021 and DHT22 sensors (respectively) in both Arduino and Raspberry Pi shields.

The following scheme shows the data structure for each air quality data block:
  
  \begin{lstlisting}[firstnumber=1]
        ["1570884025": {
              "temperature": 22.5,
              "humidity": 56.05,
              "co2": 430.3,
              "co": 5,
              "so2": 55,
              "no2": 20,
              "pm10": 14.2,
              "pm2.5": 9,
              "bateryLevel": "95%",
              "hash": "2456615909b3bd..."
              ...
        }]
\end{lstlisting}

\begin{table}[]
	\centering
	\caption{Specifications of integrated environmental sensors}
	\label{tab:environmental_specification}
	\begin{tabular}{|c|c|c|c|c|c|}
	    \hline
		Variable    &   Success Rate    &   Read            &   Power Supply    &   Operating Range     &   Conditions\\ \hline
		T           &   DHT22           &   \  Digital \    &   3.3-6 V       &   T:-40-80ºC          &   -\\ \hline
		H           &   DHT22           &   \  Digital \    &   3.3-6 V       &   H:0-100\%          &   -\\ \hline
        CO          &   MQ-7            &   \  Analog \     &   5 V             &   20-2000 ppm       &   T:-20\textcelsius -50\textcelsius,H:\textless95\%  \\ \hline
        CO2         &   TGS4161         &   \  Analog \     &   5 V             &   350-10000 ppm     &   T:-10\textcelsius$\sim$50\textcelsius,H:5$\sim$95\%  \\ \hline
        NO2         &   MiCS-2714       &   \  Analog \     &   4.9-5.1 V     &   0.5-10 ppm        &   T:-30\textcelsius -85\textcelsius,H:5-95\% \\ \hline
        SO2         &   2SH12          &   \  Analog \     &   5 V             &   1-200 ppm         &   - \\ \hline
        Smoke       &   MQ-2            &   \  Analog \     &   5 V             &   300-10000 ppm     &   T:-20\textcelsius -50\textcelsius,H:\textless95\% \\ \hline
        PM          &   SDS021          &   \  Analog \     &   4.7-5.3 V     &   0-9999 $\mu$ g /m3  &   - \\ \hline
	\end{tabular}
\end{table}

In order to perform analogue readings using the aforementioned gas sensors, pinout or channels of analogue type are used and the Equation \ref{n_voltage} is required to convert the raw data initially obtained. To this respect, raw data goes from 0 to 1023, so it is necessary to turn it into a voltage value from 0.0 - 5.0 V (volts). This normalized voltage value is used to obtain an approximate gas concentration on the basis of the datasheet corresponding to the particular gas sensor:

\begin{equation}
\label{n_voltage}
    normalized\_voltage = raw\_data * (0.5 /1023,0)
\end{equation}

Several limits have been established to define the environmental status of every monitored polluting gas, and to generate environmental notifications and alerts. The measurement thresholds harmful to the health of people have been extracted and contrasted from different official sources of Spain \cite{airquality:Online} \cite{indoorairquality:Online}  \cite{indoormadrid:Online}. These limits are represented in Table \ref{tab:treshold} for the average exposure for each variable with regard to the exposure time for each one.

\begin{table}[]
	\centering
	\caption{Thresholds of pollutant gases}
	\label{tab:treshold}
	\begin{tabular}{|c|c|c|c|c|c|c|}
		\hline
		Variable                    &   Average                     &   Time                 \\ \hline
		CO2                         &   500-1000 ppm                &   -                    \\ \hline
		CO                          &   10 $\mu$g/m3                &   year                 \\ \hline
		\multirow{2}{*}{NO2}        &   200 $\mu$g/m3 &  hour \\    &   40 $\mu$g/m3  & year \\ \hline
		\multirow{2}{*}{SO2}        &   350 $\mu$g/m3 &  hour \\    &   125 $\mu$g/m3 & day  \\ \hline
		PM$_{2.5}$                  &   25 $\mu$g/m3                &   year                 \\ \hline
		\multirow{2}{*}{PM$_{10}$}  &   50 $\mu$g/m3 &  day \\      &   40 $\mu$g/m3 & year  \\ \hline

	\end{tabular}
\end{table}

\section{Applied method}
\label{sec:method}

Several statistical calculations are applied to the environmental measurements registered by every IoT device that composes a distributed WSN. Each environmental measurement package includes the concentration of each pollutant gas considered (CO2, CO, NO2, SO2 and smoke), particles (PM2.5 and PM10) and, finally, temperature and humidity values. For each new measurement of each variable included in a new air quality data block, the system calculates the following fields, taking into account a set denoted as \(N\) that involves the last 50 measurements registered:

\begin{enumerate}
	\item The new measurement that corresponds to the last value collected for a particular monitored environmental variable.
	\item Average of \(N\) measurements.
	\item Assignment of class mark to the last measurement value. For this purpose, the rule of Sturges \cite{scott2009sturges} is initially applied on the basis of \(N = 50\) last measurements of each variable (see Equation \ref{sturges}). Thus, \(K \) classes are considered and the corresponding class range for each one is  calculated according to the measurement thresholds of the associated sensor. Finally, the new measurement is classified depending on its value. Likewise, absolute and relative frequencies are  calculated on the basis of \(N = 50\) last measurements.
	\begin{equation}
	\label{sturges}
	    K = 1 + log_{2}(N), A = \frac{max. threshold - min. threshold}{K}
	\end{equation}
	\item Variance (\(s^{2}\)) and standard deviation (\(s\)) are according to Equation \ref{variance}. These statistical values allow the system to determine the level of the representation of the average with regard to \(N\) environmental values. The aim is to determine the average distance of each measurement of the \(N\) set with respect to the previous calculated average value. Both the web application and the mobile version allow users access to these values, so that they can compare in real time the dispersion of the polluting gas measurements or environmental measurement within each monitored indoor area (temperature and humidity) through different graph types.
	\begin{equation}
	\label{variance}
	    s^{2} = \frac{\sum_{i=1}^{n} (x_{i} - \bar{x})^{2}}{n-1}, s = \sqrt{\frac{\sum_{i=1}^{n}(x_{i}-\bar{x})^{2}}{n-1}}
	\end{equation}
	\item Coefficient of variation (\(c.v.\)). Once the standard deviation has been calculated, the relative variability of the last \(N\) measurements registered is studied (Equation \ref{coef_variation}). The result is expressed in percentages.
	\begin{equation}
	\label{coef_variation}
	    c.v. = \frac{s}{\bar{x}}
	\end{equation}
\end{enumerate}

These calculations can be used to monitor air quality throughout the space under consideration. With regard to this, the trend of change, dispersion and variability of each pollutant gas, temperature and humidity across the indoor areas are delimited and monitored by each IoT device. Thus, through an analysis of the effect of the existing factors on the environment of each monitored area, a dependency relationship could be established between the variability of the variable and the causative factors.

Once the proposed statistical calculations are obtained for each variable included in a new air data quality block, the system updates the implemented database and allows users to access to this environmental data through integrating interactive linear and bar graphs. To this respect, users can compare the current state and the variability of a particular gas variable across several indoor zones Thus, they can analyse standard deviation, variation, frequency of certain range measurements,  average and other aforementioned statistical calculations. 


\section{Security of the proposal}
\label{sec:security}

Different methods have been applied with the aim of ensuring the secure processing of air quality data measured by distributed IoT devices. The main purpose is the development of mechanisms capable of securing the authenticity, confidentiality and integrity of measured environmental information. Particular attention has been paid to the implementation of a robust authentication method for system users that sign in by the mobile application and the Web service.

\subsection{Security scheme}

\begin{figure}
	\centering
	\includegraphics[width=\textwidth]{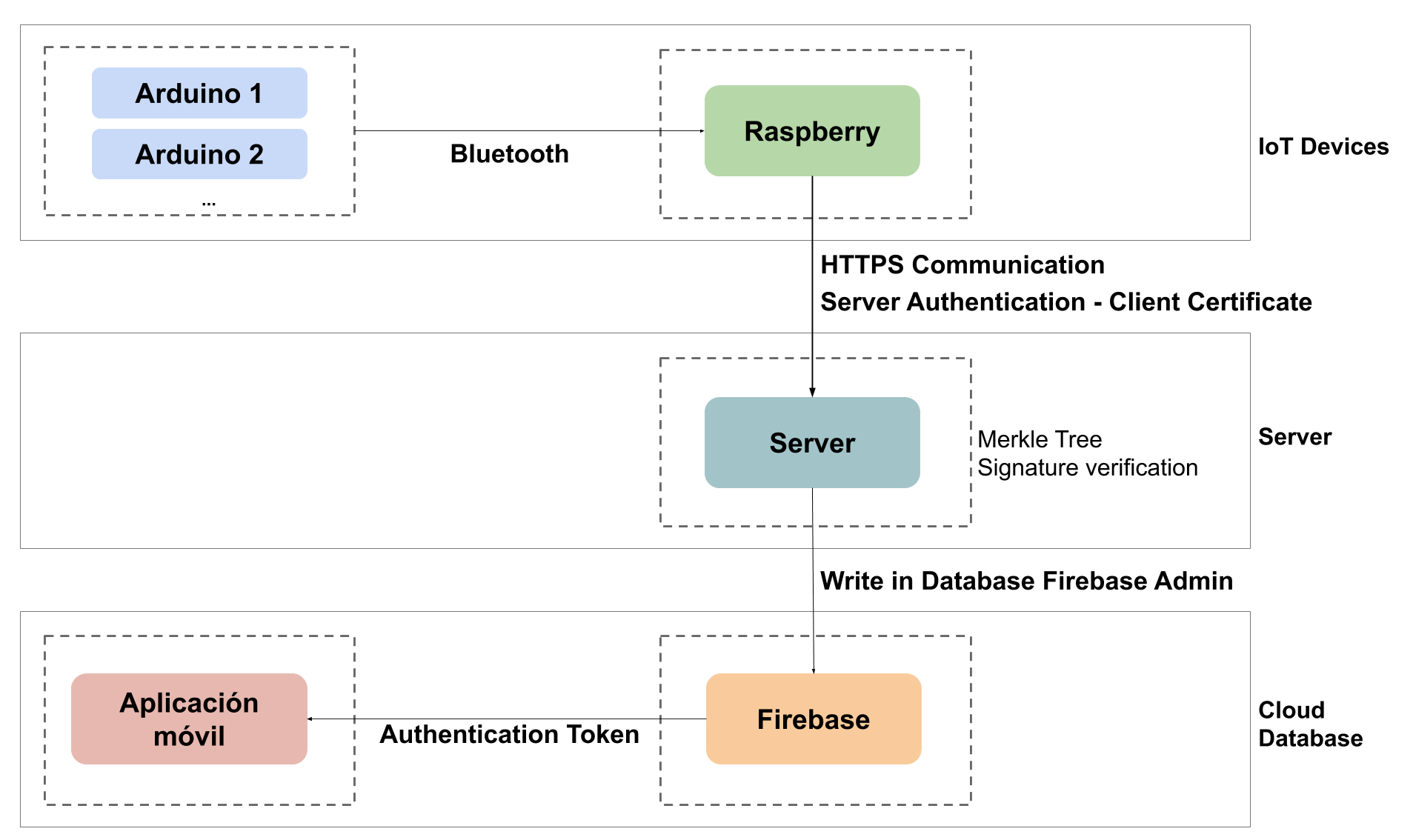}
	\caption{Security Scheme}
	\label{fig:security_scheme}
\end{figure}

Figure \ref{fig:security_scheme} shows the proposed security scheme. Different aspects have been considered with the aim of providing security mechanisms to authenticate sensor nodes or IoT devices.

Regarding the authentication of sensor nodes, each IoT device connects to the Cloud server via an HTTPS connection through providing in each request its client certificate X.509 \cite{cooper2008internet}. This certificate is validated by the implemented server in order to obtain the IoT device identifier. Once the Merkle tree integrity is ensured, the new air quality data block received is stored in the database. Different security rules have been configured  to add new air quality data blocks in the implemented Firebase database. To this respect, the Firebase admin role is in charge of updating air quality information instead of directly allowing  database writing performed by IoT devices.

In order to sign in the mobile application and the Web application, a custom authentication token is required to authenticate a user against the implemented cloud server and Firebase. This token is generated from the user's UID via RS256 (RSA Signature with SHA-256) and includes several configuration parameters, including an expiration time. When this token expires, the user's session is cancelled and the system forces the user to log in again (both in the web application and in the mobile application). In addition, other security guidelines have been applied in the system following Open Web Application Security Project (OWASP) guidelines (such as the encryption of integrated API keys like Google Maps or Firebase, strong password generation policies, etc.) \cite{owasp:Online}.

\subsection{Merkle Tree}
\label{merkletree}

Merkle trees or hash trees are a tree-shaped data structure used to verify the integrity of large amounts of data blocks. To this respect, they have been used to verify the integrity of the air quality measurements registered by the IoT devices, so allowing their transfer in networks without corruption or alteration. For each measurement cycle, every sensor node measures the current concentration of different pollutant gases (such as CO2, CO, SO2, NOx, etc.) and the value of temperature and humidity to finally build a new air quality data block. Each air quality data block is processed and digested by applying a hash  function, aimed at obtaining a new leaf node of the implemented Merkle tree. To this respect, leaf nodes can be linked to a unique root hash value contained in the root node of the tree. This root node provides the integrity of the full tree that involves all air quality information blocks measured and added to the tree structure. Any small change made to a particular data block causes changes in the rest of the non-leaf nodes of the hash tree, and thus also modifying the root hash. Thus, these changes can be easily detected by verifying the value of the root node hash for every new air quality block measured.

\begin{figure}
	\centering
	\includegraphics[width=\textwidth]{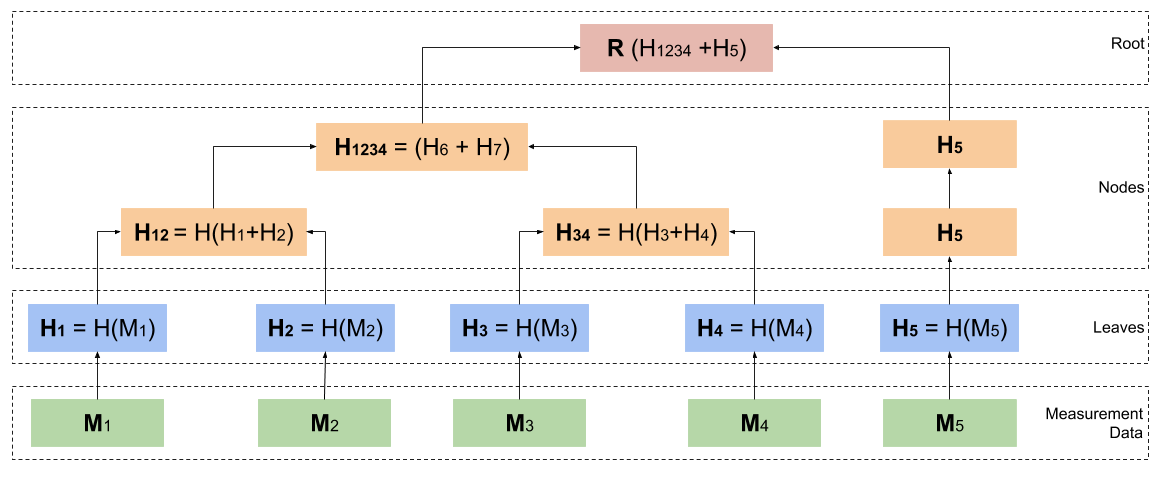}
	\caption{Merkle Scheme}
	\label{fig:merkle_scheme}
\end{figure}

The result of applying the SHA-3 hash cryptography function \cite{dang2015secure} to the value of each new block of measurements generated provides a new leaf node that is added to the proposed Merkle tree. The introduction of new leaves in the tree finally modifies its structure, the content of the non-leaf nodes of the higher levels that structure it and the value of root hash. As Figure \ref{fig:merkle_scheme} shows, every tree node contains the result of applying the SHA-3 hash function to its pair of children nodes, and so following a binary tree structure. As the equation \ref{eq:sha3} shows, every air quality data block denoted as Mx produces a value of SHA-3 digest represented as H (Mx) (leaf node). This value is concatenated with another SHA-3 value H(Mx) that corresponds to another air quality data block (leaf node), so the hash of concatenated result constitutes their parent non-leaf node. This process is successively executed to create the root of the tree at the highest level.

\begin{equation}\label{eq:sha3}
H_{12} = H(H_{1} | H_{2}) = H (H(M_{1})| H(M_{2}))
\end{equation}

Equation \ref{eq:nodes_dimen} determines the growth of the implemented Merkle tree, which means the increase in the number of tree levels as a consequence of adding new leaf nodes. In this regard, for a number of leaves \(h\) inserted in the tree, there is a variable \( n\) that also belongs to the set of natural numbers \({0,1,2,3,4, ...}\), so satisfying  \(2^{(n-1)} < h \leq 2^{n}\) and providing a tree dimension \(size (tree)\) according to \(ln(2^{n})/ln(2) + 1\).

\begin{equation}
\label{eq:nodes_dimen}
    \forall h \epsilon \mathbb{N},\\
    \exists n \epsilon \mathbb{N} \mid 2^{n-1} <  h \leq 2^{n} \Rightarrow Size(tree) = Ln(2^{n})/ Ln(2) + 1
\end{equation}

\begin{equation}
\label{eq:nodes_num}
Number\_of\_nodes = 2L - 1
\end{equation}

Table \ref{tab:nodes_growth} shows the growth of the implemented tree and the changes produced in its structure from the treatment of the first 12 air quality data blocks registered by an IoT device. For each environmental data block,  a new leaf is inserted in the tree by applying SHA-3 on data block content. To this respect, as a consequence of inserting each new leaf in the tree structure, each row shows the number of  tree levels and the total number of nodes generated (including leaf nodes, non-leaf nodes and root node). The number of nodes for \(L\) leaf nodes added to the Merkle tree is determined according to Equation \ref{eq:nodes_num}.

\begin{table}[]
	\centering
	\caption{Threshold pollutant gases}
	\label{tab:nodes_growth}
	\begin{tabular}{|c|c|c|c|}
		\hline
		Leaves number   &  \ \(2^n\) \  &   Tree Level                  &   Number of Nodes                 \\  \hline
		                &               &   \ \(ln(2^n) / ln(2)+1\) \   &   \ \(2 * num(leaves) - 1 \) \    \\  \hline
		1               &   1           &   1                           &   1                               \\  \hline
		2               &   2           &   2                           &   3                               \\  \hline
		3               &   4           &   3                           &   5                               \\  \hline
		4               &   4           &   3                           &   7                               \\  \hline
		5               &   8           &   4                           &   9                               \\  \hline
		6               &   8           &   4                           &   11                              \\  \hline
		7               &   8           &   4                           &   13                              \\  \hline
		8               &   8           &   4                           &   15                              \\  \hline
		9               &   16          &   5                           &   17                              \\  \hline
		10              &   16          &   5                           &   19                              \\  \hline
		11              &   16          &   5                           &   21                              \\  \hline
		12              &   16          &   5                           &   23                              \\  \hline
	\end{tabular}
\end{table}

\begin{figure}
	\centering
	\includegraphics[width=\textwidth]{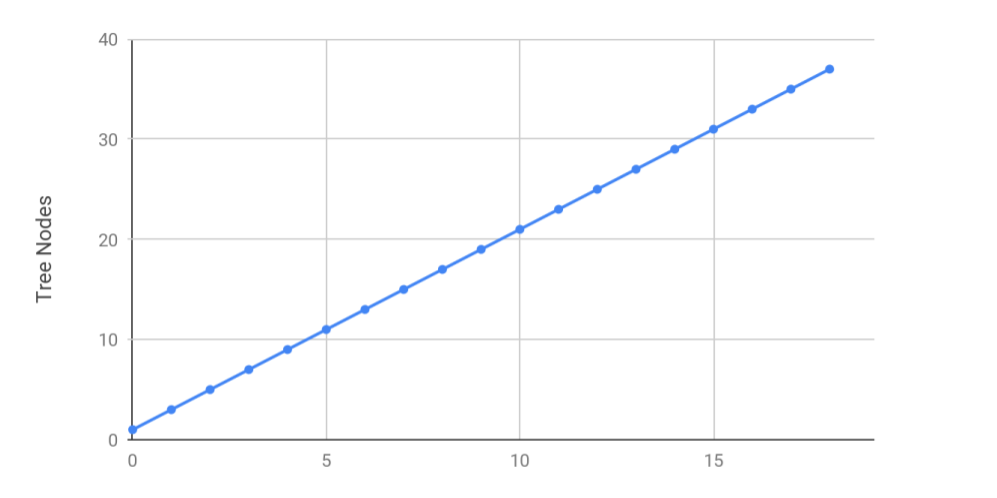}
	\caption{Number of Merkle tree nodes}
	\label{fig:treenodesgraphs}
\end{figure}

The following Figure \ref{fig:graphs} shows the relationship between the number of leaves (air quality data blocks) inserted into the Merkle tree. Thus, the increase in Merkle tree levels with regard to the increase in leaf node introduction according to the Equation \ref{eq:nodes_dimen} is detailed. On the other hand, Figure \ref{fig:treenodesgraphs} represents the number of tree nodes generated as a consequence. In this case, this graph shows the comparison of the number of existing tree nodes (including leaf-nodes, non-leaf nodes and the root) with respect to the current number of leaf nodes (according to the Equation \ref{eq:nodes_num}).

\begin{figure}
	\centering
	\includegraphics[width=\textwidth]{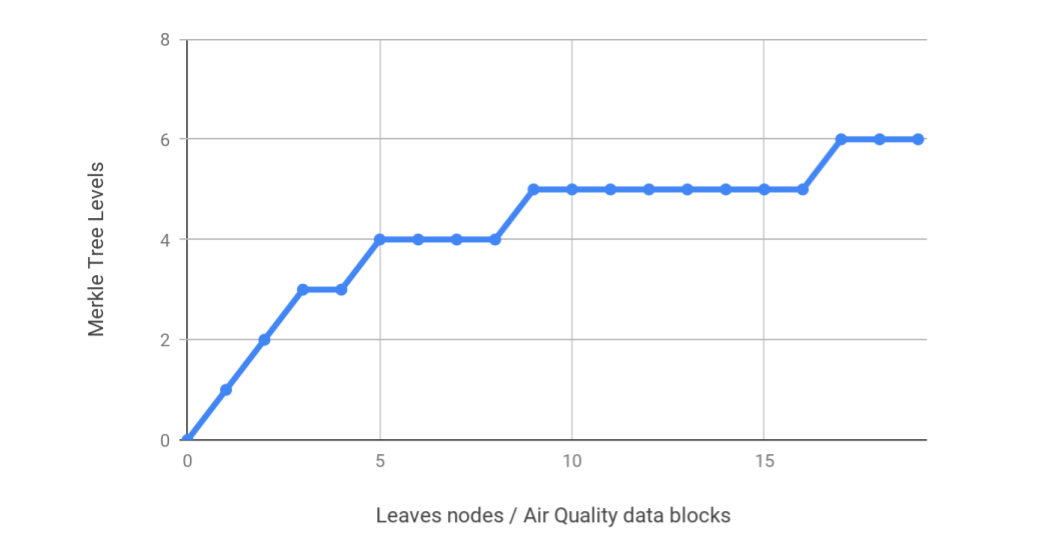}
	\caption{Calculating Merkle Tree levels}
	\label{fig:graphs}
\end{figure}

The server must check that a new block or package of environmental measurements collected by an IoT device belongs to the implemented Merkle tree. For this purpose, its integrity must be verified using the following parameters that are sent by the IoT device:

\begin{enumerate}
	\item Source data. Environmental measurements registered by the IoT device that constitute a new block of air quality information.
	\item The hash values of non-leaf intermediate nodes required to obtain the new root node. To verify the integrity of the tree and the newly inserted leaf node, it is not necessary to provide the original data from other environmental data blocks previously registered by the same IoT device.
	\item The hash value of the new proposed root node that has been recently modified by inserting the new data block in the tree structure, with regard to the implemented Merkle tree in the IoT device controller that has recently measured source air quality data.
\end{enumerate}

In order to verify the integrity of the new content of the Merkle tree, the server applies the same SHA-3 hash function to the source data (1) to get the value of the new leaf node to be inserted in the tree. The result obtained is concatenated with the hash values of the non-leaf intermediate nodes received (2) and digested by applying SHA-3. Finally, the server obtains the value of the root hash according to the data received from the IoT device. This root value is compared with that received from the IoT device (3) with the aim of verifying the Merkle tree and ensuring the integrity of air quality data. To this respect, it is possible to check if the new air quality data block is truly part of the data in the Merkle tree and verify if the air quality measurements have not been modified and corrupted. The server updates the implemented Merkle tree and performs the signature of the new root whose result is finally sent to the IoT device. On the other hand, the new air quality data block is disregarded when the verification process fails, so a security notification is automatically sent to the system administrator.

\begin{table}[]
\caption{Data Blocks}
\label{tab:data_blocks}
\begin{tabular}{ll}
{[}"828...": \{ "t": 22.0, "h": 56.05, "co2": 430.3, ......, "bat": "95\%", "hash": "24566.."\}{]}  \\
{[}"379...": \{ "t": 21.5, "h": 56.39, "co2": 441.2, ......, "bat": "95\%", "hash": "0a961..."\}{]} \\
{[}"998...": \{ "t": 21.8, "h": 55.19, "co2": 445.2, ......, "bat": "92\%", "hash": "cf3d7..."\}{]} \\
{[}"492...": \{ "t": 22.0, "h": 54.03, "co2": 453.5, ......, "bat": "90\%", "hash": "122cd..."\}{]} \\
{[}"226...": \{ "t": 22.5, "h": 55.50, "co2": 454.1, ......, "bat": "90\%", "hash": "55848..."\}{]} \\
{[}"848...": \{ "t": 22.7, "h": 58.12, "co2": 445.3, ......, "bat": "85\%", "hash": "6f674..."\}{]} \\
{[}"235...": \{ "t": 22.5, "h": 58.83, "co2": 436.7, ......, "bat": "88\%", "hash": "0d29c..."\}{]} \\
{[}"323...": \{ "t": 22.2, "h": 60.09, "co2": 439.9, ......, "bat": "90\%", "hash": "776ae..."\}{]} \\
{[}"549...": \{ "t": 21.0, "h": 62.50, "co2": 435.3, ......, "bat": "89\%", "hash": "29ea0..."\}{]} \\
{[}"737...": \{ "t": 20.5, "h": 62.29, "co2": 432.1, ......, "bat": "85\%", "hash": "7f271..."\}{]}
\end{tabular}
\end{table}

Table \ref{tab:data_blocks} shows a dataset composed of 10 examples of air quality data blocks measured by an Arduino-based sensor node of the implemented WSN. Each line of the table represents a data block based on the parameters and variables explained in the proposed method in this work. 

With regard to this, Table \ref{tab:hash_merkle_blocks} shows the Merkle tree obtained by digesting the value of each air quality data block through applying the SHA-3 hash function. Thus, the corresponding 10 leaves are obtained and used to build the tree with the aim of establishing a root node that ensures the tree integrity. In this table, it is possible to analyse the growth and changes applied to the Merkle tree on the basis of adding its leaf nodes. Likewise, the result of successively concatenating these nodes is shown in order to  obtain the corresponding parent node or root hash. Thus, the root node of the implemented Merkle tree is updated. This process is presented in detail in the following example shown in Table \ref{tab:hash_merkle_blocks}.

\begin{table}[]
\caption{Merkle tree nodes}
\label{tab:hash_merkle_blocks}
\begin{tabular}{{|c|c|c|c|c|}}
\hline
Hash Data      & Level 2                           & Level 3                           & Level 4                           & Root                                  \\ \hline
2456615909b3...     & \multirow{2}{*}{ f542a5459a35...} & \multirow{4}{*}{ 05956bc36b92...} & \multirow{8}{*}{ aa953c896cf2...} & \multirow{10}{*}{ 57e7bf9e937f...}    \\ \cline{1-1}
0a9618d27355...     &                                   &                                   &                                   &                                       \\ \cline{1-2}
cf3d77b82f74...     & \multirow{2}{*}{ 43b3918b6410...} &                                   &                                   &                                       \\ \cline{1-1}
122cd92d6178...     &                                   &                                   &                                   &                                       \\ \cline{1-3}
5584804be30b...     & \multirow{2}{*}{ 0f041e1fb933...} & \multirow{4}{*}{ f7d1c5a1757b...} &                                   &                                       \\ \cline{1-1}
6f674266f36f...     &                                   &                                   &                                   &                                       \\ \cline{1-2}
0d29cf6b8714...     & \multirow{2}{*}{ 16400da437...}   &                                   &                                   &                                       \\ \cline{1-1}
776ae014e8aa...     &                                   &                                   &                                   &                                       \\ \cline{1-4}
29ea0d5b0884...     & \multirow{2}{*}{ 56ac4a1b66f0...} & \multirow{2}{*}{ 56ac4a1b66f0...} & \multirow{2}{*}{ 56ac4a1b66f0...} &                                       \\ \cline{1-1}
7f2715599417...     &                                   &                                   &                                   &                                       \\ \hline
\end{tabular}
\end{table}

\section{Conclusions}
\label{sec:conclusions}

This paper proposes a system to monitor and analyse indoor air quality in real time, specifically focused on its appliance at critical transport infrastructure. Indoor air quality depends on several ambient factors and the activities that are being carried out, so it is essential to monitor it with regard to this type of infrastructures. Innovative technologies have been used, such as the deployment of distributed wireless sensor networks with IoT devices. As a consequence, the implemented system provides indoor air pollution alerts when polluting gas concentrations unusually increase, so deteriorating the air quality level as a result of several factors, such as intense human traffic. To this respect, air pollution could cause problems for people's health. The proposed system  implements a Web service and a mobile application that allow users the visualization and analysis of air quality measurements registered by a particular IoT prototype in real time. The Raspberry Pi and Arduino platforms have been used to assemble gas sensors and other environmental sensors aimed at constituting  each node of a distributed Wireless Sensor Network.

Particular attention has been paid to the implementation of security methods to manage the system information. Merkle trees are proposed as a solution to protect the integrity of the air quality data sent by each node, with the SHA-3 algorithm being applied to every new information block. This method guarantees the integrity of each new package with regard to the previous leaf nodes existing in the tree. To this respect, each request from the IoT node to the server is made on the basis of an HTTPS connection by attaching the corresponding client certificate. Regarding the authentication of system users, authentication tokens are applied to access the system and the registered environmental data. These user tokens are generated from the identifier of each user by means of the signature RS256 (RSA signature 256).

Currently, work is in progress on the application of Fuzzy Classification Trees in this system. Regarding possible improvement in the analysis of air quality data blocks, a hierarchical structure is being developed through decision trees in a diffuse environment, in order to improve the interpretation of inaccurate information.

\section*{Acknowledgements}
Research supported by the Spanish Ministry of Science, Innovation and Universities, the  FEDER Fund, the Centre for the Development of Industrial Technology and the CajaCanarias Foundation, under Projects RTI2018-097263-B-I00, C2017/3-9, IDI-20160465 and DIG02-INSITU.

%
%
\bibliography{bibliography.bib}
\bibliographystyle{spbasic}

\end{document}